\documentclass[prl,twocolumn,amsmath,amssymb,showpacs,noshowkeys,superscriptaddress]{revtex4}
\usepackage{graphicx}
\usepackage{dcolumn}
\usepackage{bm}
\usepackage{color}

\begin{document}

\title{Multiple Changes of Order of the Vortex Melting Transition in $\mathrm{Bi_2Sr_2CaCu_2O_8}$ with Dilute Columnar Defects}

\author{T.~Verdene}
\email{talhazak@wisemail.weizmann.ac.il}
\author{H.~Beidenkopf}
\author{Y.~Myasoedov}
\author{H.~Shtrikman}
\author{M.~Rappaport}
\author{E.~Zeldov}
\affiliation{Department of Condensed Matter Physics, Weizmann
Institute of Science, Rehovot 76100, Israel}%
\author{T.~Tamegai}
\affiliation{Department of Applied Physics, The University of Tokyo, Hongo, Bunkyo-ku, Tokyo 113-8656, Japan}%
\date{\today}

\begin{abstract}
A low concentration of columnar defects (CDs) is reported to
transform a first-order vortex lattice melting line in
$\mathrm{Bi_2Sr_2CaCu_2O_8}$ crystals into alternating segments of
first- and second-order transitions separated by two critical
points. As the density of CDs is increased, the critical points
shift apart and the range of the intermediate second-order
transition expands. A third, low temperature critical point was
also observed in one sample. The measurement of equilibrium
magnetization and the mapping of the melting line down to 27K was
made possible by employment of the shaking technique.
\end{abstract}
 \keywords{} \maketitle

The interplay between different energy scales of a physical system
can induce phase transitions and yield a rich phase diagram. In
particular, in a clean system a melting transition results from
the competition between elasticity and thermal fluctuations and is
usually of a first-order nature. The presence of a third competing
scale due to disorder can suppress it to second order (SO) and
lead to an even richer picture, the understanding of which is
still quite limited \cite{Imry, Fisher}. In order to gain insight
into the various competing mechanisms, we introduce an additional
control parameter in the form of a variable weak correlated
disorder. We report an exceptionally intricate case, where the
initially first-order (FO) melting transition of the vortex matter
in a high temperature superconductor exhibits a rare three-section
FO-SO-FO behavior and even seems to undergo a four-section
FO-SO-FO-SO sequence. Our findings reveal that a SO transition is
nucleated in a small segment along the FO transition line, bound
by two critical points. This SO segment grows when the control
parameter is increased thereby providing important insight into
the general mechanism of transformation of a FO transition into a
SO transition.

$\mathrm{Bi_2Sr_2CaCu_2O_8}$ (BSCCO) high-$T_c$ superconductor has
an especially rich vortex matter phase diagram which results from
the interplay between the elasticity, thermal fluctuations, and
point disorder. It consists of a FO melting line \cite{Eli Nature}
which separates the low-field Bragg glass from the high-field
disordered phases \cite{giamarchi, Nattermann, {Blatter}}. At high
temperatures the FO melting transition is mainly thermally driven,
whereas at lower temperatures it gradually changes its nature into
a disorder-driven FO inverse melting transition \cite{Nurit,
{Brandt_PD}, {Teitel}}. Indications of an additional almost
vertical SO glass transition line at intermediate temperatures
were also found \cite{Haim1}.

This phase diagram changes entirely with introduction of a fourth
energy scale due to dense columnar defects (CDs) $B_{\phi}\gg
B_m(T)$, where $B_{\phi}\equiv n_{cd}\phi_0$, $n_{cd}$ is the
density of CDs, $\phi_0$ is the flux quantum, and $B_m(T)$ is the
temperature-dependent melting field. Compared to the correlated
pinning energy, the elastic energy becomes negligible. As a
result, the ordered solid phase is replaced by a disordered Bose
glass (BoG) phase in which vortices are localized on CDs and melt
through a SO BoG transition \cite{Nelson_and_Vinokur, {Blatter}}.

In this paper we focus on the dilute CD limit, where vortices
outnumber CDs at most relevant fields. Remarkably, in this limit
the energy due to correlated pinning is comparable to the other
three energy scales. Consequently, the four relevant energy scales
mold together a particularly complex $B-T$ phase diagram whose
nature is not well-understood. Most experimental \cite{Khaykovich,
{Satya1}, {Menghini}, {Konczy}, {Nurit2}}, theoretical
\cite{Radzihovsky, {Rodriguez}, {Larkin&vinokur}} and numerical
studies \cite{{Goldschmidt}, {Nonomura},{DasGupta}} agree that at
high temperatures the transition remains FO. At intermediate
temperatures where the melting occurs at intermediate fields
$B_m\gtrsim B_{\phi}$, the pristine Bragg glass phase is replaced
by a porous solid with the majority of CDs occupied by strongly
pinned vortices that form an amorphous matrix. The remaining
vortices form ordered crystallites that are embeded in the pores
of the rigid  matrix \cite{Satya1, {Menghini}, Radzihovsky,
{Larkin&vinokur}, {Goldschmidt}, {Rodriguez}, {Nonomura},
{DasGupta}, {Nurit2}}. These vortex nanocrystals melt into
nanodroplets apparently through a second order transition.

 Experimentally, even a low density of CDs
enhances hysteretic effects throughout most of the solid phase in
BSCCO \cite{Khaykovich, {Konczy}}. This is a major obstacle in
mapping the true thermodynamic phase transitions of vortex matter.
Lacking access to the thermodynamic behavior, past experimental
studies have focused on dynamics  \cite{{Konczy}, {Marcin},
Khaykovich,{Satya1}, {Nurit2}}. It was concluded that the high
temperature FO transition line terminates at a critical point (CP)
whose location depends on CD density \cite{Khaykovich, {Satya1},
{Menghini}, {Nurit2}}, consistent with theory and simulations
\cite{Larkin&vinokur, {Radzihovsky}, {Goldschmidt}, {Rodriguez},
{Nonomura}, {DasGupta}}. At temperatures below the CP, however,
these out-of-equilibrium studies could not evaluate the exact
nature of the transition. Thus, a SO transition was premised to
exist at all temperatures below the CP \cite{Khaykovich, {Satya1},
{Radzihovsky}, {Goldschmidt}, {Rodriguez}, {Nonomura},
{DasGupta}}, though other theoretical scenarios were also
considered \cite{Larkin&vinokur}. This incomplete understanding of
the dilute pinning limit led naturally to the assumption that as
CD density is increased, the SO nature of melting spreads from low
temperatures to higher ones with a shift of a single CP. Our
findings, however, suggest an essentially different process.

Within the present study we mapped for the first time the
\textit{equilibrium} phase diagram of vortex matter with dilute
CDs by performing local magnetization measurements during vortex
shaking. This method \cite{Willemin, {Haim1}, {Nurit}, {Brandt}}
utilizes an in-plane ac shaking field which agitates the vortices
and assists them to assume their equilibrium positions. Tilting
the magnetic field away from the CDs was shown to alter the
equilibrium magnetization in YBCO crystals \cite{Hayani}. In
contrast, in BSCCO moderate in-plane field has essentially no
effect on the equilibrium properties even in presence of CDs due
to the very high anisotropy \cite{drost, Nurit2}. As a result, the
ac in-plane field enhances vortex relaxation without altering the
thermodynamic transitions \cite{Nurit2}.
 We
measured three BSCCO samples irradiated at GANIL with matching
fields of $B_{\phi}=\mathrm{5G, 10G, 20G}$ and critical
temperature $T_c\approx 92$K, and an additional sample with
$B_{\phi}=\mathrm{20G}$ and $T_c\approx 91$K. Half of each sample
was masked during irradiation to allow a direct comparison between
pristine and irradiated behavior. The local magnetic induction of
the sample was measured by an array of Hall sensors 10x10 $\mu
$m$^2$ each, while sweeping the external magnetic field $H$ at a
constant temperature. Simultaneous measurements of the transition
on the pristine halves of the samples show that it remained FO
throughout the temperature range. Shaking fields up to $350$Oe at
$\mathrm{10Hz}$ were used.

\begin{figure} [!t] \centering
\mbox{\includegraphics[width=0.49\textwidth]{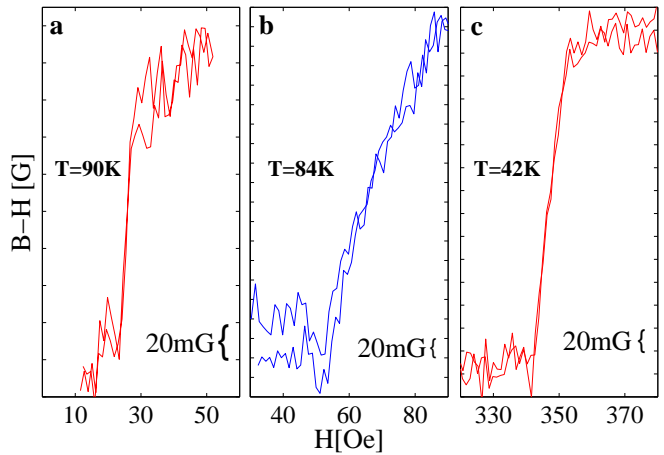}}
\caption{(color online) The measured local magnetization, $B-H$,
of the $B_{\phi}=\mathrm{10G}$ sample in the presence of shaking,
measured upon ascending and descending external field at constant
temperatures of (a) $\mathrm{90K}$, (b) $\mathrm{84K}$ and (c)
$\mathrm{42K}$. A linear slope was subtracted for clarity. The
sharp steps in the local magnetization in (a) and (c) signify a FO
transition. In (b), a break in the slope signifies a SO
transition.} \label{measuremetns}
\end{figure}

\begin{figure} [!b] \centering
\mbox{\includegraphics[width=0.43\textwidth]{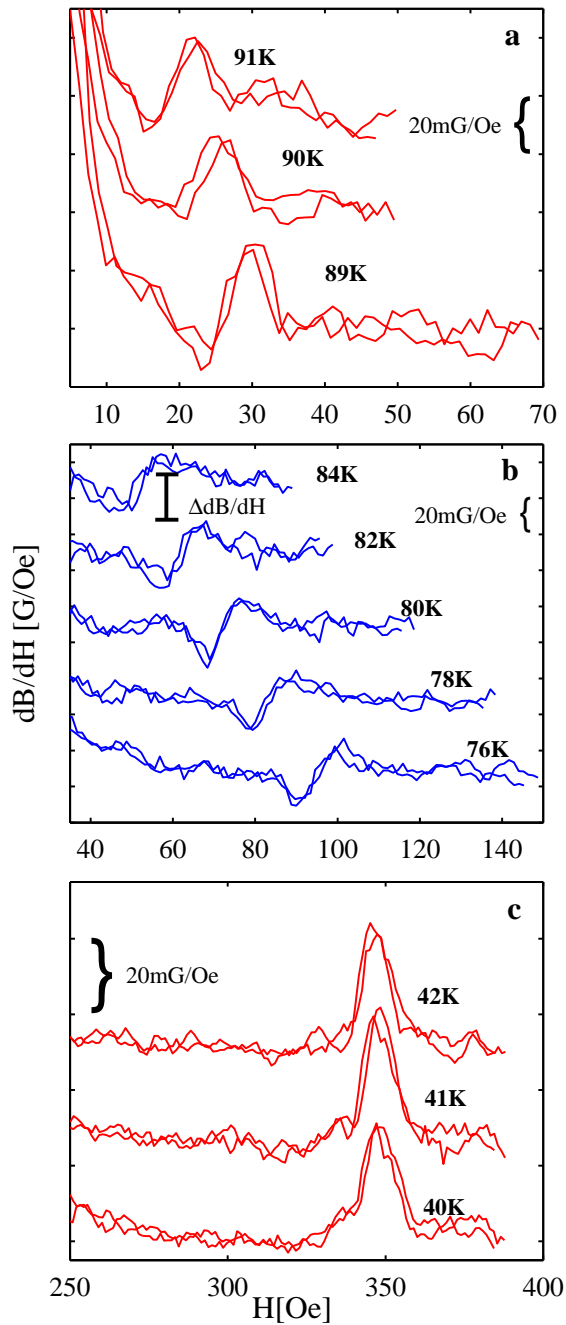}}
\caption{(color online) The derivative of the measured induction
with respect to applied field, $dB/dH$, of the
$B_{\phi}=\mathrm{10G}$ sample in presence of shaking, measured
upon ascending and descending field at constant temperatures of
(a) $\mathrm{89K-91K}$, (b) $\mathrm{76K-84K}$ and (c)
$\mathrm{40K-42K}$. Data are shifted vertically for clarity. A
peak in $dB/dH$ in (a) and (c) signifies a FO transition, whereas
a step in (b) indicates a SO transition.} \label{derivatives}
\end{figure}

 Figure \ref{measuremetns} presents the
measured local magnetization, $B-H$, at (a) 90K, (b) 84K and (c)
42K. A FO transition appears as a sharp step in $B(H)$ and a SO
transition is manifested by a break in its slope. To better
resolve the nature and location of the phase transition we
differentiated the measured induction $B$ with respect to the
applied field $H$. Figure \ref{derivatives} shows the derivatives
$dB/dH$ of representative (a) high temperature, (b) intermediate
temperature and (c) low temperature measurements. A
 $\delta$-like peak in the derivative $dB/dH$ indicates a FO transition, whereas a discontinuity
 signifies a SO transition.

At high temperatures we find that the transition remains FO in the
presence of CDs, similarly to that in the pristine areas, as
previously reported \cite{Satya1, {Khaykovich}}. Accordingly,
$B(H)$ in Fig. \ref{measuremetns}a shows a sharp step and the
derivatives $dB/dH$ in Fig. \ref{derivatives}a display
$\delta$-like peaks.
At a slightly lower temperature thermal fluctuations are weaker,
the effective pinning potential of the CDs gains strength and
irreversibility is greatly enhanced, thus masking the underlying
phase transition. We overcame this problem by applying the shaking
method rendering a reversible magnetization \cite{Willemin,
{Brandt}, Haim1, Nurit}. Figures \ref{measuremetns}b and
\ref{derivatives}b thus show equilibrium measurements at
intermediate temperature of $B-H$ and $dB/dH$, respectively.
 A break in the
slope of the induction is clearly resolved along both ascending
and descending field sweeps and the derivatives display a step
structure indicating a SO transition.

Yet, our main finding is the recovery of the FO thermodynamic
phase transition at lower temperatures. This is clearly visible in
Fig. \ref{measuremetns}c as a reversible discontinuity in
magnetization $B(H)$ and accordingly, in Fig. \ref{derivatives}c a
sharp peak in the derivatives $dB/dH$. This recurrence of the FO
nature of the transition has been predicted theoretically
\cite{Larkin&vinokur}, but has never been observed experimentally.

These findings can be explained by the following comparison of
energy scales. At high temperatures thermal fluctuations are
dominant enough to weaken the effective pinning potential due to
CDs \cite{{Larkin&vinokur}, {Radzihovsky}, {Goldschmidt},
{Rodriguez}, {Nonomura}, {Menghini}, {DasGupta}, {Khaykovich},
{Satya1}, {Nurit2}, colson}, resulting in a FO transition similar
to that found in pristine samples. At intermediate temperatures,
as correlated pinning gains dominance, the randomly distributed
CDs alter the \textit{equilibrium} vortex matter state, in
addition to enhancing irreversible hysteretic behavior. At these
temperatures the melting occurs at intermediate fields $B_m\gtrsim
B_{\phi}$ and the pristine Bragg glass phase is replaced by a
porous solid with crystallites of interstitial vortices imbedded
in the pores of a rigid amorphous vortex matrix \cite{Satya1,
{Menghini}, {Nurit2}}.
 The size of the nanocrystals within each pore is only
several times their lattice constant. These dilute nanocrystals
seem to melt into nanodroplets through a SO transition probably
since the range of correlations is cutoff by the finite size of
the pores, consistent with recent numerical simulations
\cite{Goldschmidt, {Rodriguez}, {Nonomura}, {DasGupta}}. This is
similar to the dense CD limit, where correlated pinning dominates
and vortex matter undergoes a second order BoG transition.

\begin{figure} [!b] \centering
\mbox{\includegraphics[width=0.49\textwidth]{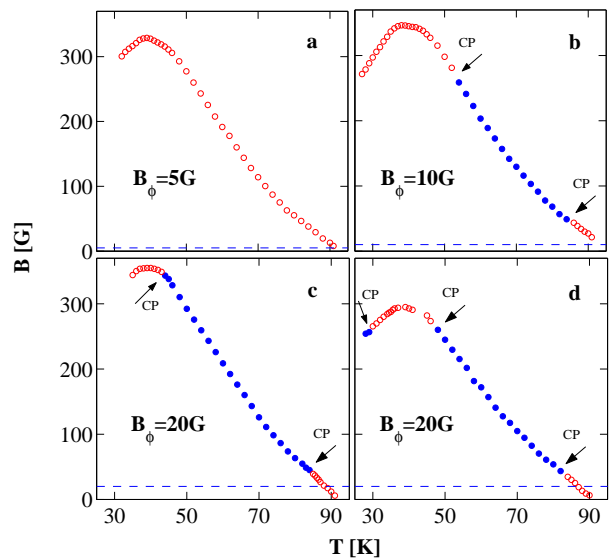}}
\caption{(color online) Phase diagrams of irradiated BSCCO
crystals showing FO (\textcolor[rgb]{0.94,0.00,0.00}{$\circ$}) and
SO (\textcolor[rgb]{0.00,0.00,1.00}{$\bullet$}) transitions. (a)
In $B_{\phi}=$5G sample the transition is FO at all temperatures.
(b) For $B_{\phi}=$10G a FO-SO-FO behavior is found. (c) The SO
segment expands in $B_{\phi}=$20G sample. (d) A different
$B_{\phi}=$20G sample with lower doping that exhibits a
FO-SO-FO-SO sequence. The values of $B_{\phi}$ are shown by dashed
lines. \label{phase_diagrams}}
\end{figure}

At low temperatures $\sim40$K the transition occurs at high
fields. Each pinned vortex is surrounded by tens of interstitial
vortices and the solid phase is increasingly dominated by inter
vortex interactions \cite{Larkin&vinokur}, resulting in relatively
ordered dense vortex crystallites \cite{Menghini}. Consequently,
their melting into nanodroplets is accompanied by
 a discontinuity in the entropy that reflects the sharp
difference in ordering of the two phases. The FO nature of the
transition is thus restored, as in pristine samples.

Note that at intermediate temperatures, where the transition is
SO, the step in the derivative is positive, $\Delta \partial
B/\partial H>0 $ (see Fig. \ref{derivatives}b). This is consistent
with a positive step in the specific heat, $\Delta C_v>0$ which
was found in the high-field SO transition of
pristine YBCO \cite{Bouquet, {Schilling}}. It is in contrast,
however, to the negative step of the derivative, $\Delta
\partial B/\partial T<0$, at the SO glass transition reported in pristine BSCCO \cite{Haim1}.
Further study is required in order to explain the origin of these
differing signs.

Interestingly, at intermediate temperatures $dB/dH$ takes a more
intricate shape; rather than a simple upward jump, it first
decreases, then jumps up and finally slightly decreases again.
This structure may be explained by the association of the step in
the derivative at the SO transition with the relative ordering of
the two phases across the transition line. Then, the decrease in
the derivative might be interpreted as a slight ordering of
interstitial vortices prior to the major disordering upon melting.
A similar feature has been observed in the structure factors in
Monte-Carlo simulations \cite{Goldschmidt}. It is yet unclear,
however, how these possible structural changes should affect the
equilibrium magnetization.

The location and nature of the transition of the four samples was
measured and mapped. Shaking enabled the detection of a melting
transition down to $\mathrm{27K}$. The results are summarized in
Fig. \ref{phase_diagrams}. The FO and SO regions are marked by
open and solid circles respectively. In all four irradiated
samples, the location of the melting line remained similar to that
of the pristine parts of the samples, with a maximum at
$\sim\mathrm{40K}$. The $\mathrm{5G}$ sample exhibits a FO
transition at all temperatures. As the density of CDs is
increased, more of the $B-T$ phase space becomes dominated by
correlated disorder and consequently, a larger portion of the
transition line becomes SO. The $\mathrm{10G}$ and $\mathrm{20G}$
samples both display two CPs with a FO-SO-FO sequence. In the
$\mathrm{20G}$ sample the two CPs are shifted further apart than
in the $\mathrm{10G}$ sample and the SO transition spreads both to
high and low temperatures. We therefore suggest that this trend
persists until a full SO BoG transition line is attained in the
dense-pin limit.

Note that the FO transition line of irradiated samples persists to
the left of its maximum in the inverse melting region (Fig.
\ref{phase_diagrams}). These are the first equilibrium
magnetization measurements in this region in presence of CDs.
Figure \ref{phase_diagrams}d shows the phase diagram of a sample
with $B_{\phi}=\mathrm{20G}$ and a slightly lower oxygen doping.
This sample displays even more intricate behavior. Like the other
$B_{\phi}=\mathrm{20G}$ and $\mathrm{10G}$ samples, it also
exhibits  FO-SO-FO behavior. In addition, this sample reveals a
third CP at extremely low temperatures on the inverse melting side
of the transition. Below this CP we observe a recurrence of the SO
nature of the transition. It can be attributed to the fact that
the melting field decreases in the inverse melting region with
decreasing T. As a result, nanocrystals contain less vortices and
correlated pinning regains dominance over elasticity.
Consequently, the transition becomes SO once again, similarly to
that at intermediate temperatures.

It is worth pointing out the qualitative difference between the
behavior reported here and the SO-FO-SO sequence found in pristine
samples of the less anisotropic YBCO compound \cite{Schilling}. In
irradiated BSCCO all three portions of the transition line
separate a solid phase from a liquid phase. In pristine YBCO,
however, the high-field SO portion is believed to separate two
liquid phases due to the existence of a tri-critical point
\cite{Bouquet}. The low field SO portion, on the other hand,
separates a solid phase from a liquid phase and is believed to
arise from the intrinsic correlated disorder in YBCO crystals
\cite{Welp, {Kwok}}.

In summary, we present thermodynamic evidence for new FO-SO-FO
behavior of the melting line in BSCCO samples irradiated with a
low density of CDs. This unusual behavior is due to close
competition between four different energy scales in the dilute CD
limit. As CDs are introduced, the melting transition initially
alters its order from first to second at intermediate temperatures
where correlated pinning has a dominant effect. At high and low
temperatures the transition remains FO despite the disordering
potential of CDs. At low temperatures where the melting field is
high, the solid phase is dominated by elasticity rather than
correlated pinning due to the increasing number of vortices per
CD, and the FO transition is retained. The FO nature of the
transition is likewise preserved at high temperatures, where
correlated pinning is weakened by thermal fluctuations. As the
density of the CDs is increased, the SO segment of the transition
line expands both to the higher and lower temperatures. In one
sample a SO transition was also found on the low-temperature side
of the inverse melting portion, resulting in an even more complex
FO-SO-FO-SO sequence. The observed nucleation and growth process
of the SO segments along the original FO transition line clarifies
the general process of transformation of phase transitions with
increased disorder. In particular, it describes the mechanism that
leads to transformation of a Bragg glass in the presence of point
disorder to a BoG at high concentrations of correlated disorder,
which melts through a single SO transition.

We thank C.J.~van der Beek and M.~Konczykowski for sample
irradiation and Y.Y. Goldschmidt, E.H. Brandt and G.P. Mikitik for
useful discussions. This work was supported by the German Israeli
Foundation (GIF), US-Israel BSF, and Grant-in-aid from the
Ministry of Education, Culture, Sport, Science and Technology,
Japan.


\begin{thebibliography}{10}
\bibitem{Imry} Y. Imry and M. Wortis,  Phys. Rev. B \textbf{19},
3580 (1979).
\bibitem{Fisher} D.S. Fisher,  M.P.A. Fisher and D.A. Huse,  Phys. Rev. B \textbf{43},
130 (1991).
\bibitem{Eli Nature} E. Zeldov \textit{et al.},  Nature \textbf{375},
373 (1995).
\bibitem{giamarchi}
T.~Giamarchi and P.~LeDoussal,
\newblock Phys. Rev. B {\bf 52}, 1242 (1995).
\bibitem{Nattermann} T. Nattermann,  Phys. Rev. Lett. \textbf{64},
2454 (1990).
\bibitem{Blatter} G. Blatter and V.B. Geshkenbein, \textit{The Physics of Superconductors} (Springer, New York, 2003).
\bibitem{Nurit} N. Avraham \textit{et al.},  Nature \textbf{411},
451 (2001).
\bibitem{Brandt_PD}G.P. Mikitik and E.H. Brandt, Phys. Rev. B \textbf{68},
054509 (2003).
\bibitem{Teitel} P. Olsson and S. Teitel,  Phys. Rev. Lett. \textbf{87},
137001 (2001).
\bibitem{Haim1} H. Beidenkopf \textit{et al.},  Phys. Rev. Lett. \textbf{95},
257004 (2005).
\bibitem{Nelson_and_Vinokur} D.R. Nelson and V.M. Vinokur,  Phys. Rev. Lett. \textbf{68},
2398 (1992).
\bibitem{Konczy} M. Konczykowski \textit{et al.}, Physica C {\bf 408},
547 (2004).
\bibitem{Khaykovich} B. Khaykovich \textit{et al.},  Phys. Rev. B \textbf{57},
14088 (1998);\newblock B. Khaykovich \textit{et al.},
\newblock Physica C {\bf 282}, 2068 (1997).
\bibitem{Satya1} S.S. Banerjee \textit{et al.},  Phys. Rev. Lett. \textbf{90},
087004 (2003);\newblock S.S. Banerjee \textit{et al.},  Phys. Rev.
Lett. \textbf{93}, 097002 (2004).
\bibitem{Menghini} M. Menghini \textit{et al.},  Phys. Rev. Lett. \textbf{90},
147001 (2003).
\bibitem{Nurit2} N. Avraham \textit{et al.},  Phys. Rev. Lett. \textbf{99},
087001 (2007).
\bibitem{Larkin&vinokur} A.I. Larkin and V.M. Vinokur,  Phys. Rev. Lett. \textbf{75},
4666 (1995).
\bibitem{Radzihovsky} L. Radzihovsky,  Phys. Rev. Lett. \textbf{74},
4923 (1995).
\bibitem{Rodriguez} J. P. Rodriguez, Phys. Rev. B \textbf{70}, 224507 (2004).
\bibitem{Goldschmidt} S. Tyagi and Y. Y. Goldschmidt,  Phys. Rev. B \textbf{67},
214501 (2003);\newblock Y. Y. Goldschmidt and E. Cuansing, Phys.
Rev. Lett. \textbf{95}, 177004 (2005); \newblock Y. Y. Goldschmidt
and Jin-Tao Liu, Phys. Rev. B \textbf{76}, 174508 (2007).
\bibitem{Nonomura} Y. Nonomura and X. Hu, Europhys. Lett. \textbf{65}, 533 (2004).
\bibitem{DasGupta} C. Dasgupta \textit{et al.},
Phys. Rev. B \textbf{72}, 94501 (2005).
\bibitem{Marcin} C.J. van der Beek \textit{et al.},  Phys. Rev. Lett. \textbf{86},
5136 (2001).
\bibitem{Willemin} M. Willemin \textit{et al.},  Phys. Rev. Lett. \textbf{81},
4236 (1998).
\bibitem{Brandt} G.P. Mikitik and E.H. Brandt,  Phys. Rev. B \textbf{69},
134521 (2004);\newblock Superconductor Science and Technology
\textbf{20}, 9 (2007)
\bibitem{Hayani} B. Hayani \textit{et al.},  Phys. Rev. B \textbf{61},
717 (2000).
\bibitem{drost} R.J. Drost \textit{et al.},  Phys. Rev. B \textbf{58},
615 (1998).
\bibitem{colson} S. Colson \textit{et al.},  Phys. Rev. B \textbf{69},
180510 (2004).
\bibitem{Schilling} A. Schilling \textit{et al.},  Phys. Rev. Lett. \textbf{78},
4833 (1997).
\bibitem{Bouquet} F. Bouquet \textit{et al.},  Nature \textbf{411},
448 (2001).
\bibitem{Welp} U.Welp \textit{et al.},  Phys. Rev. Lett. \textbf{76},
4809 (1996).
\bibitem{Kwok} W.K. Kwok \textit{et al.},  Phys. Rev. Lett. \textbf{84},
3706 (2000).
\end{thebibliography}
\end{document}